\documentclass[a4paper,fleqn]{cas-dc}
\pdfoutput=1
\usepackage[numbers]{natbib} 
\usepackage{algorithm}
\usepackage{algpseudocode}
\usepackage{graphicx}
\usepackage{textcomp}
\usepackage{hyperref}
\usepackage{xcolor}
\usepackage{booktabs}
\usepackage{multirow}
\usepackage{cleveref}
\usepackage{makecell}
\usepackage{subcaption}

\usepackage{todonotes}

\definecolor{codegreen}{rgb}{0,0.6,0}
\definecolor{codegray}{rgb}{0.5,0.5,0.5}
\definecolor{codepurple}{rgb}{0.58,0,0.82}
\definecolor{backcolour}{rgb}{0.95,0.95,0.95}

\definecolor{delim}{RGB}{20,105,176}
\definecolor{numb}{RGB}{106, 109, 32}
\definecolor{string}{rgb}{0.64,0.08,0.08}

\begin{document}
\let\WriteBookmarks\relax
\def\floatpagepagefraction{1}
\def\textpagefraction{.001}

\shorttitle{CloudSim 7G: An Integrated Toolkit for Modeling and Simulation of Future Generation Cloud Computing Environments}

\shortauthors{R. Andreoli, J. Zhao, T. Cucinotta, and R. Buyya}

\title [mode = title]{CloudSim 7G: An Integrated Toolkit for Modeling and Simulation of Future Generation Cloud Computing Environments}                    

\author[1,2]{Remo Andreoli}
\cormark[1]
\ead{remo.andreoli@santannapisa.it}

\affiliation[1]{organization={CLOUDS Lab},
            addressline={
School of Computing and Information Systems, University of Melbourne}, 
            city={Melbourne},
            country={Australia}}

\author[1]{Jie Zhao}

\author[2]{Tommaso Cucinotta}

\author[1]{Rajkumar Buyya}

\affiliation[2]{organization={ReTiS Lab},
            addressline={ TECIP Institute, Sant'Anna School of Advanced Studies}, 
            city={Pisa},
            country={Italy}}
            
\cortext[cor1]{Corresponding author}

\begin{abstract}
Cloud Computing has established itself as an efficient and cost-effective paradigm for the execution of web-based applications, and scientific workloads, that need elasticity and on-demand scalability capabilities. However, the evaluation of novel resource provisioning and management techniques is a major challenge due to the complexity of large-scale data centers. Therefore, Cloud simulators are an essential tool for academic and industrial researchers, to investigate the effectiveness of novel algorithms and mechanisms in large-scale scenarios.
This paper proposes CloudSim 7G, the seventh generation of CloudSim, which features a re-engineered and generalized internal architecture to facilitate the integration of multiple CloudSim extensions within the same simulated environment.   As part of the new design, we introduced a set of standardized interfaces to abstract common functionalities and carried out extensive refactoring and refinement of the codebase. The result is a substantial reduction in lines of code with no loss in functionality, significant improvements in run-time performance and memory efficiency (up to 25\% less heap memory allocated), as well as increased flexibility, ease-of-use, and extensibility of the framework. These improvements benefit not only CloudSim developers but also researchers and practitioners using the framework for modeling and simulating next-generation Cloud Computing environments.
\end{abstract}


\ExplSyntaxOn
\keys_set:nn { stm / mktitle } { nologo }
\ExplSyntaxOff
\maketitle

\section{Introduction}
Over the past decade, Cloud Computing has evolved rapidly, becoming the dominant model for modern computing. Compared to previous paradigms, the characterizing aspects of Cloud Computing are the illusion of infinite resources available 24/7 on-demand over the network, and of infinitely scalable applications. This is possible thanks to virtualization technologies, which allow the sharing of physical machines, storage, and networking devices among multiple customers and organizations. A plethora of service providers are involved in the provisioning of Cloud services, including Cloud and Network Service providers, to enable a wide range of applications being built by customers. The goal of the Cloud provider is to ensure a satisfying experience for the customer, usually in terms of performance, reliability, security, and cost, without compromising profit. 

It is difficult and costly to evaluate policies for Cloud provisioning, workload management, and resource handling in a real environment since a Cloud infrastructure is a large and complex system comprising interconnected geo-located datacenters.
In this context, simulation toolkits~\cite{Markus20,mansouri2020cloudSimSurvey,Bambrik20} mitigate the issue and play an important role in research communities for testing and evaluating complex applications and novel resource management strategies through inexpensive and repeatable experiments. Furthermore, simulations provide a controlled environment to test the performance of resource provisioning policies and to easily reproduce the results. 

CloudSim~\cite{calheiros2011cloudsim} has been a forefront simulation toolkit, and the de-facto standard, for evaluating resource management techniques in Cloud Computing environments, thanks to its ease-of-use and extensibility.
The first iteration of CloudSim offered a machine virtualization layer to simulate Virtual Machines (VMs) and test techniques such as provisioning, scheduling, and consolidation. Thousands of contributions in the field of Cloud orchestration and resource management use CloudSim~\cite{awan2024dfarm, satpathy2023GAMap,wu2011SLAbasedResAlloc}, as demonstrated by the 6400+ citations to the seminal paper. Indeed CloudSim is a completely customizable tool: all its components and related interactions are implemented in Java and can be extended with minimal effort. For instance, Cloud researchers can develop a custom scheduling policy, or embed power-awareness in their VM instances. As the number and type of services offered by Cloud service providers have increased, in parallel, simulation toolkits have also evolved. As a result, CloudSim sports a rich ecosystem of extensions, from now on referred to as ``CloudSim modules'', to model and simulate all sorts of resource management challenges in the Cloud context. The rapid growth in Cloud adoption not only relies on machine virtualization as its main feature, but also on many fundamental features such as Software Defined Networking (SDN), Network Function Virtualization (NFV), container-based virtualization, and serverless application execution models. The CloudSim research community developed new modules such as NetworkCloudSim~\cite{garg2011networkcloudsim}, CloudSimSDN~\cite{son2015cloudsimsdn, son2019cloudsimsdnnfv}, WorkflowSim~\cite{chen2012workflowsim}, and ContainerCloudSim~\cite{piraghaj2017containercloudsim}, among others, to accommodate these advancements.

Since its inception, the CloudSim ecosystem has received contributions from diverse researchers and developers with varying skill levels and coding styles. As highlighted by other authors~\cite{Sukhoroslov2022dslab,silva2017cloudsimplus, mastenbroek2021opendc}, these contributions often fail to fully leverage the toolkit's extensibility and are typically developed as independent packages, significantly hampering the reuse of
these extensions. These ``single-feature'' extensions led to a degree of fragmentation within the ecosystem, making it difficult for CloudSim users to seamlessly integrate multiple modules into a single simulated scenario.
The culprit is CloudSim's lack of generalized interfaces for implementing the simulated entities and their interactions. A module developer creating an entirely new component had no easy way to maintain compatibility with components from other, unrelated CloudSim modules: developers have often been forced to copy-paste existing components and slightly modify their behaviors, to the detriment of the extensibility capabilities offered by Object-Oriented (OO) programming. For instance, ContainerCloudSim provided a collection of container scheduling and allocation policies which were the copy-pasted version of the VM-based ones. This is because VMs and containers have been portrayed as completely different components, despite performing the same task (i.e., executing user workloads). Another example: ContainerCloudSim provided a collection of host selection policies for placement purposes, whereas the power-aware package offered a collection of VM selection policies for migration purposes. The policies are fundamentally the same in both modules (i.e., select an entity from a list of candidates), but CloudSim does not provide a standard way to interface with them, forcing redundant code paths: one to handle migration, and one to handle placement. Moreover, the latest version of CloudSim (i.e., CloudSim 6G) was released with a bundle of these independently packaged modules with no interoperability guarantees, further worsening the readability and clarity of the codebase.

\subsection{Contributions}\label{subsec:contrib}
This paper proposes the seventh generation of the CloudSim toolkit, shortened to CloudSim 7G, the biggest re-engineering of the codebase to date. The major contributions of our proposal are:
\begin{enumerate}
    \item A set of Java interfaces to standardize the definition, configuration, and creation of new components and their interactions. Compared to original CloudSim~\cite{calheiros2011cloudsim}, these interfaces facilitate the integration of several CloudSim modules, which were previously available independently and often had compatibility issues, within the same simulated scenario.

    \item A re-engineered, refactored, and refined version of several modules from CloudSim 6G, for a total of more than 13,000 lines of code removed. In particular: i) NetworkCloudSim~\cite{garg2011networkcloudsim}, the networking module of CloudSim, has been rewritten almost completely; and ii) ContainerCloudSim~\cite{piraghaj2017containercloudsim}, which is the container module, and the power module~\cite{beloglazov2013dynamicConsolidation} have been heavily refactored to reduce their codebase and improve readability as well as extensibility.

    \item The ability to simulate scenarios with nested virtualization, such as containers within VMs, or even VMs within VMs. This enables a more accurate simulation of real-world Cloud infrastructures, where nested virtualization is used to implement multi-tenancy and enhance isolation.

    \item The introduction of the virtualization overhead parameter: in real-world settings, VMs do not communicate directly with physical devices, such as network interface cards, but rather through virtual interfaces, which introduce additional overhead~\cite{chen2015virtOverhead}. This feature can be used to characterize better scenarios involving nested virtualization and networking. 
        
    \item Significant improvements in terms of simulation run-time performance and memory usage, compared to previous CloudSim iterations.
\end{enumerate}

The third and fourth contributions are made possible by the standardized interfaces introduced in CloudSim 7G.
The pre-packaged modules available in CloudSim 6G, which were unable to work in an integrated manner, have been consolidated into a \textit{base layer} to support the development of novel multi-module extensions. Moreover, significant attention has been given to minimizing the effort required for module developers to restore compatibility of their contributions with CloudSim 7G.

In conclusion, CloudSim 7G introduces enhanced functionalities and extensibility features, along with significant performance improvements.
These innovative capabilities open up new opportunities for simulating next-generation Cloud Computing environments, where researchers can experiment with hybrid scenarios using various computing paradigms (i.e., Cloud, Fog, Edge, Serverless Computing) and deployment models (i.e., VMs only, containers only, containers on VMs, etc.) together. This ability for different modules to coexist and interact was previously not possible.

\textbf{\subsection{Paper Organisation}\label{subsec:organisation}}
The rest of the paper is organized as follows: \Cref{sec:cloudsim_evol} recalls the evolution of the official CloudSim codebase, as released by researchers of the CLOUDS Lab, in conjunction with the paradigm shifts in Cloud Computing. \Cref{sec:relwork} briefly presents a collection of external modules not developed within the CLOUDS Lab, as well as a subset of CloudSim's competitors in the literature. \Cref{sec:cloudsim7g} presents: i) the proposed architectural change to the codebase of CloudSim so to allow for multi-extension simulations; ii) the essential refactoring, refinement, and optimization steps performed to construct the new base layer of CloudSim; and iii) the guidelines to update an old module to CloudSim 7G.
\Cref{sec:performance_eval} presents a performance comparison between CloudSim 6G and CloudSim 7G in terms of run-time and total memory allocated, showcasing the improvements of CloudSim 7G in large simulations thanks to the massive optimization the codebase has undergone. \Cref{sec:case-study} proposes a simple case study that uses multiple modules simultaneously, demonstrating the flexibility of the novel design.  Finally, \Cref{sec:conclusions} concludes the paper with final remarks and a discussion on future research directions for modeling and simulating future generation Cloud Computing environments. 

\section{CloudSim Through the Evolution of Cloud Computing and Related Paradigms}\label{sec:cloudsim_evol}
\begin{figure*}
    \centering
    \includegraphics[width=0.9\linewidth]{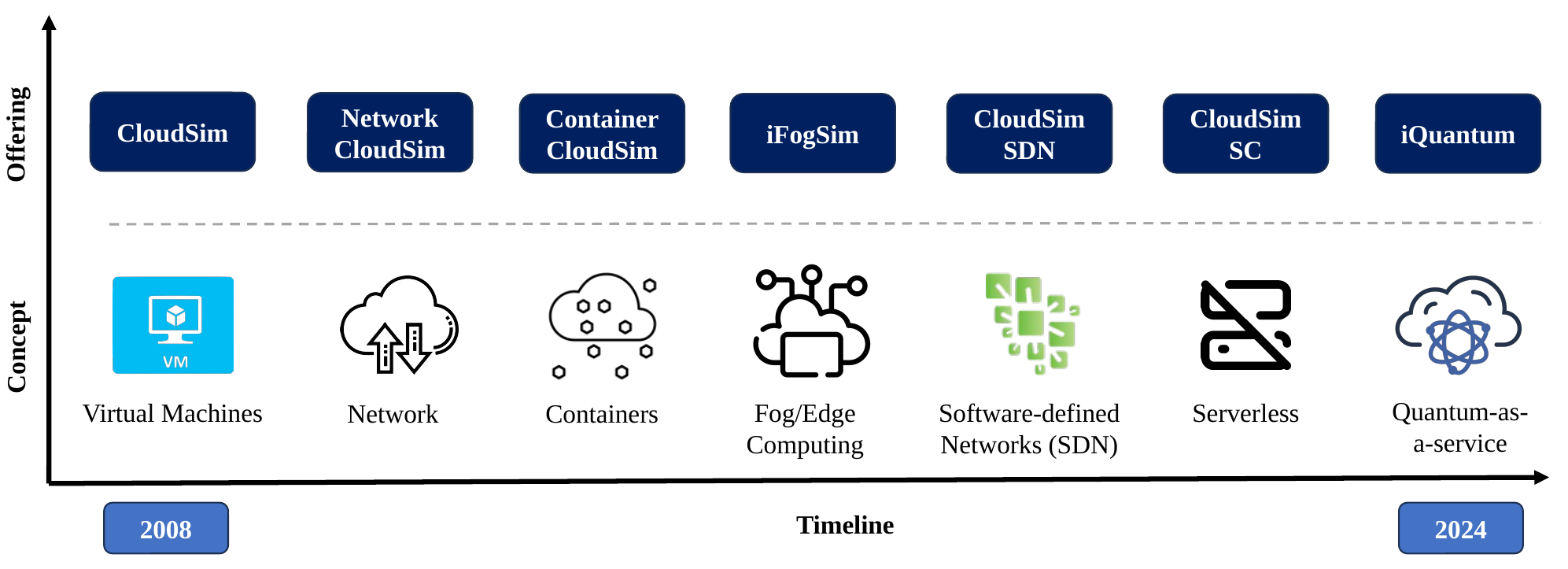}
    \caption{The evolution of CloudSim}
    \label{fig:cloudsim-evolution}
\end{figure*}

In the landscape of Cloud simulators available in the literature~\cite{Markus20,mansouri2020cloudSimSurvey,Bambrik20}, CloudSim, one of the first simulators specialized in evaluating resource management techniques for Cloud infrastructures, has established itself as the de-facto choice for the research community.
This section recalls the major characteristics of the milestones in the evolution of CloudSim, depicted in \Cref{fig:cloudsim-evolution}, as officially released by researchers at the CLOUDS Lab.
Notice that some of the described modules have been integrated into the base layer of CloudSim 7G.

CloudSim originated as a derivative of GridSim~\cite{buyya2002gridsim}, enhancing its core functionalities with a new discrete-event management framework that more effectively represents the dynamic nature of Cloud environments. The first iteration of CloudSim~\cite{buyya2009cloudsim, calheiros2011cloudsim} modeled the basic functionalities of machine virtualization described above, simple network behaviors in the form of a constant propagation delay calculated using a latency matrix, and a power module for investigating energy-aware VM consolidation policies.
One of the shortcomings of the initial version of CloudSim is the absence of a proper network model. This limitation precludes the modeling of realistic scenarios with datacenter network topologies~\cite{liu2013dcn}, which is crucial for a range of Cloud application domains, including scientific, big-data, and High-Performance Computing (HPC) workloads. In these contexts, applications are typically composed of several communicating tasks that need to be distributed across a cluster of interconnected physical machines.
NetworkCloudSim~\cite{garg2011networkcloudsim} addressed such limitations with the introduction of a network flow model that simulates an internal network made of interconnected switches and physical machines. Consequently, NetworkCloudSim also introduced a generalized application model for simulating complex workflow applications using a message-passing paradigm for communications. A ``networked'' cloudlet is structured as a sequence of stages, where each stage represents either a computational activity, like a ``traditional'' cloudlet, or a data transmission dependency (i.e., send/receive data). 
The communication between cloudlets that are part of the same workflow application but reside on different physical hosts is managed through a series of simulated switches.

The next evolution step came with the advent of Containers-as-a-Service (CaaS) offerings in the Cloud market. Containers represent a progression towards lightweight application management~\cite{pahl2019containers}, hence it became evident the need for simulating such a novel virtualization paradigm.  To this end, ContainerCloudSim~\cite{piraghaj2017containercloudsim} enriched CloudSim with new features for the evaluation of allocation, migration, and scheduling policies in containerized environments. In practice, ContainerCloudSim adds an additional scheduling layer to the simulated infrastructure, so that cloudlets are deployed within containers that are placed inside VMs.

The widespread adoption of virtualization technologies eventually reached the telecommunications field with the rising interest in Network Function Virtualization (NFV). Together with Software-Defined Network (SDN) technologies, they enable the deployment of network services, such as proxy and firewall ones, on commodity hardware instead of proprietary, expensive hardware appliances, much like virtualization is employed in Cloud infrastructures to drive higher capacity utilization and reduce cost. CloudSimSDN~\cite{son2015cloudsimsdn, son2019cloudsimsdnnfv} enabled the evaluation of resource management strategies in infrastructures with SDN functionalities, such as dynamic network configuration and programmable controllers. In addition, CloudSimSDN supports the allocation, migration, autoscaling, and service chaining for NFV. The simulation framework is based on the Open Source NFV Management and Orchestration (MANO) architecture~\cite{ersue2013etsi}. 

The advent of the Internet-of-things (IoT) paradigm created the need for pushing computation and storage away from centralized data centers, and towards the ``edge'' of the network, to support the surge of big data. In this regard, iFogSim~\cite{gupta2017ifogsim} enabled the simulation of IoT devices connected to Edge and Fog Computing environments. The latest version~\cite{mahmud2022ifogsim2} supports also the creation of complex microservice applications, as well as the simulation of service migrations for different mobility models of IoT devices and distributed cluster formation among Edge/Fog nodes of different hierarchical tiers.

The subsequent step in the history of CloudSim was supporting the evaluation of cloud-native applications~\cite{gannon2017cloudNativeApp}. In this context, Serverless computing gained much popularity thanks to the simplification it brings to
the whole process of acquiring and managing
Cloud resources. The Cloud provider takes on full responsibility for all operational tasks for
provisioning and allocating the required resources to running applications, scheduling the applications on the infrastructure, and scaling the allocated resources to adapt to traffic changes. This enables developers to concentrate solely on the development of their cloud-native applications without the burden of server administration. CloudSimSC~\cite{mampage2023cloudsimsc} is a toolkit for modeling serverless computing environments. This contribution offers a generalized architecture for function execution, scheduling, and load balancing, as well as resource scaling and monitoring metrics in terms of application performance, system throughput, and underlying resource consumption.

The latest trends in Cloud research are investigating hybrid quantum computing environments for solving computationally intractable problems. In this regard, iQuantum~\cite{nguyen2024iQuantum} extends the base CloudSim for the evaluation of scenarios with the presence of quantum resources. The module extracts the features of quantum circuits and then models them as workload entities to be executed on quantum datacenters.

\section{Related Work}\label{sec:relwork}
The CloudSim toolkit boasts a rich ecosystem of modules developed by researchers and institutions not affiliated with the CLOUDS Lab. These external, or       ``third-party'', contributions are orthogonal to the evolution of CloudSim described in the previous subsection: sometimes an external contribution provides an alternative implementation of a milestone module or a direct improvement to it.
Finally, we discuss alternative Cloud simulators found in the literature.
\subsection{CloudSim-driven External / Third-Party Modules}\label{subsec:cloudsim-third-party-extensions}

\begin{table}
    \centering
\begin{tabular}{|c|c|c|c|}
    \hline
      Name & Based on & Use-case\\
    \hline\hline
     WorkflowSim~\cite{chen2012workflowsim}   &  CloudSim 3G  & Clustering and failure modeling for scientific workflows  \\\hline
     CloudSim4DWf~\cite{fakhfakh2017cloudsim4dwf}   &  CloudSim 4G  & Dynamic workflows  \\
    \hline
     CloudSimDisk~\cite{louis2015cloudsimdisk} & CloudSim 3G & Disk operations modeling \\\hline
     CloudSimSFC~\cite{sun2022cloudsimsfc} & CloudSim 3G &  NFV \\\hline
     ~\cite{saleh2019containerCloudSimImproved} & CloudSim 5G & Container orchestration \\\hline
     ACE~\cite{jammal2018ace} & CloudSim4G & Resiliency in Cloud \\\hline
    IoTSim~\cite{zeng2017iotsim} & CloudSim 4G & IoT for big data processing \\\hline      
     edgeCloudSim~\cite{sonmez2018edgecloudsim} & CloudSim 4G & Mobility in Edge Computing \\\hline
     IoTSimEdge~\cite{jha2020iotsimEdge} & CloudSim 5G & IoT and Edge Computing \\\hline
     CloudSimPlus~\cite{silva2017cloudsimplus} & CloudSim 3G & Re-engineering of CloudSim's core simulation framework \\\hline\hline
     CloudnetSim~\cite{cucinotta2013cloudnetsim} & OMNeT++ & Advanced CPU Scheduling \\\hline
     CloudnetSim++~\cite{malik2014cloudnetsim++} & OMNeT++ & Energy consumption \\\hline
    Simcan2Cloud~\cite{canizares2023simcan2cloud} & OMNeT++ & Machine Virtualization \\\hline
     GreenCloud~\cite{kliazovich2012greencloud} & NS2 & Energy consumption \\\hline
     DSLab IaaS~\cite{Sukhoroslov2022dslab} & DSLab & Machine virtualization\\\hline
     DSLab FaaS~\cite{semenov2024dslabFaas} & DSLab & Serverless Computing\\\hline
     FaaS-Sim~\cite{raith2023faasSim} & SimPy & Serverless in Cloud-Edge scenarios\\\hline
     OpenDC~\cite{mastenbroek2021opendc} & - & Datacenter simulation platform\\\hline
\end{tabular}
\caption{A subset of the landscape of external CloudSim modules and alternate simulators.}
\label{tab:cloud_simulators}
\end{table}

There are a plethora of external modules in the literature, those presented in this paper are summarized in \Cref{tab:cloud_simulators} (first portion). One notable contribution is
WorkflowSim~\cite{chen2012workflowsim} from the University of Southern California, an alternative to NetworkCloudSim which focuses on the simulation of the scheduling, clustering, and provisioning of large-scale scientific workflow workloads. A scientific workflow is expressed as a set of tasks that are interdependent in terms of data or control flow. In practice, the workloads are represented using the DAX format of PegasusWMS~\cite{deelman2005pegasusWMS}. Compared to NetworkCloudSim, WorkflowSim adds an additional layer of workflow management to CloudSim to evaluate workflow optimization techniques such as task clustering~\cite{topcuoglu2002heft}.  Moreover, WorkflowSim supports the generation and monitoring of failures at run-time. A more recent contribution in this context is CloudSim4WDf~\cite{fakhfakh2017cloudsim4dwf} from the University of Sfax. This extension examines the financial impact of dynamic changes in workflow. Applications are represented using Business Process Modeling Notation (BPMN) to facilitate the modeling of workload structure changes at run-time. In comparison, WorkflowSim only supports the rigid structure imposed by the DAX file.

CloudSimEx is a set of tools featuring disk operations, the simulation of MapReduce clusters~\cite{dean2008mapreduce}, web sessions, and probabilistic cloudlet arrivals, among others. CloudSimDisk~\cite{louis2015cloudsimdisk} from Luleå University of Technology is an alternative to the simple disk features of CloudSimEx that focuses on modeling the energy consumption of storage systems within a Cloud infrastructure. Compared to CloudSimEx, CloudSimDisk provides a model for describing the characteristics of a Hard Disk Drive (HDD), an integration with the power modeling package of CloudSim, and a collection of algorithms for energy-aware storage management. CloudSimSFC~\cite{sun2022cloudsimsfc}, developed by Beihang University and based on CloudSimEx, is an alternative to CloudSimSDN that enhances the simulation of performance fluctuations in service chaining within Multi-domain Service Networks. 

A contribution from German University in Cairo~\cite{saleh2019containerCloudSimImproved} provides an improved version of ContainerCloudSim that supports simulation scenarios with both VMs and containers and dynamic arrival of cloudlets and workflow applications. However, the authors re-implemented each feature from scratch. CloudSim 7G supersedes this contribution since it allows to use VMs, containers (ContainerCloudSim), workflows (NetworkCloudSim, but also WorkflowSim), and specify a dynamic cloudlet arrival (CloudSimEx) in the same scenario without ``yet another'' independently-packaged re-implementation of such features. 

FTCloudSim~\cite{zhou2015ftcloudsim} from Beijing University of Posts and Telecommunications extends CloudSim to evaluate performance in the event of fault and recovery events using checkpointing~\cite{kumari2021ftCloudSurvey}. A related work is ACE~\cite{jammal2018ace} from Western University Ontario, which investigates the resiliency of Cloud infrastructures by enhancing CloudSim with support to availability-aware placement policies. ACE allows to inject failures at VM level and recover from them through a series of recovery and repair policies. Compared to FTCloudSim, ACE also explores replication as a fault-tolerant technique.

There is a significant collection of CloudSim modules focusing on different aspects of Fog and Edge Computing. IoTSim~\cite{zeng2017iotsim} from Australian National University focuses on the simulation of IoT devices using the MapReduce model for big data processing. EdgeCloudSim~\cite{sonmez2018edgecloudsim} from Boğaziçi University is an alternative to iFogSim for modeling edge infrastructures. Compared to the first version of iFogSim, EdgeCloudSim considers mobility and implements a dynamic network communication model.  IoTSimEdge~\cite{jha2020iotsimEdge} from Newcastle University focuses on composing IoT applications as microservices and examining energy consumption implications. Compared to EdgeCloudSim, it offers enhanced mobility features.

A notable mention goes to CloudSimPlus~\cite{silva2017cloudsimplus} from Instituto Federal de Educação
Ciência e Tecnologia do Tocantins, which is proposed as an alternative to the ``base'' CloudSim. The authors present a full rewrite of CloudSim3G to reduce code duplication and increase code reuse by improving several engineering aspects of CloudSim. The goal of CloudSimPlus is in common with our work,  however, the latter suffers from two shortcomings: i) it is a fork of a very old version of CloudSim, and ii) it is not compatible with current CloudSim modules. Therefore, CloudSimPlus cannot support the rich ecosystem of contributions developed for CloudSim thus far. On the contrary, CloudSim 7G is a re-engineering attempt that stays as faithful as possible to previous CloudSim versions.

In conclusion, there is a vast array of modules that leverage the ease-of-use and extensibility of CloudSim to provide support for several computing paradigms. However, we observed significant overlap among these contributions (particularly in the context of Edge Computing): each work is an ex-novo, single-feature extension to CloudSim. This indicates that CloudSim 7G’s generalized interfaces could have facilitated building them on top of one another, rather than independently packaged modules, creating a more cohesive and readable framework.

\subsection{Other Simulators}\label{subsec:relwork}
While many contributions to the simulation and modeling of Cloud Computing are CloudSim modules, the literature also presents several alternative approaches. CloudSim remains one of the most comprehensive simulators to date, largely thanks to the extensive array of modules for experimenting with the full spectrum of Cloud-related techniques and paradigms, making it unmatched by any other Cloud simulator in terms of features. By contrast, related works tend to offer distinct advantages tailored to specific scenarios.

As highlighted by a recent survey on the topic~\cite{mansouri2020cloudSimSurvey}, most alternative Cloud simulators in the literature are extensions of existing commercial network simulators. A subset of these contributions is described in what follows and summarized in \Cref{tab:cloud_simulators} (second portion). CloudNetSim~\cite{cucinotta2013cloudnetsim} was an extension of OMNeT++~\cite{varga2010omnet++}, a commercial but open-source network simulator, to support the simulation of CPU scheduling within physical hosts and VMs, including hypervisor-level and guest OS-level hierarchical scheduling. It focused on the simulation of the Completely Fair Scheduler (CFS) default scheduler in the Linux kernel when used to schedule KVM VMs onto over-provisioned physical hosts, as well as to schedule tasks within a Linux guest VM. CloudnetSim++~\cite{malik2014cloudnetsim++} was another extension of OMNet++, developed independently from CloudNetSim, designed to study energy consumption of different components spread across geographically distributed Cloud datacenters. These two platforms exploited OMNeT++ to offer accurate simulation of the TCP/IP network protocol, but in different contexts: scheduling algorithms and energy consumption, respectively. Simcan2Cloud~\cite{canizares2023simcan2cloud} is one of the latest Cloud simulators based on OMNet++, which focuses on providing a highly-detailed 
simulated Cloud infrastructure. It models innovative aspects, including VM rental extensions, different queues for managing diverse access types, and prioritized resource allocation.

GreenCloud~\cite{kliazovich2012greencloud} was an extension of NS2, one of the most popular open-source network simulators, for evaluating energy-aware scheduling algorithms in Cloud environments. It offered detailed modeling of energy consumption for hardware resources, including networking appliances such as switches and links, with diverse compute and communication capacities, and SLA requirements. All these approaches have the advantage that they inherit packet-level accuracy from the underlying network simulator, but at the cost of long simulation times due to the large amount of ``small'' individual simulated components. Additionally, they lack support for simulating per-VM resource allocation and placement strategies, unlike a ``general-purpose'' simulator such as CloudSim.

Regarding ex-novo simulators, the authors in~\cite{Sukhoroslov2022dslab} propose DSLab IaaS, a novel simulator for IaaS infrastructures based on a general-purpose software framework for simulating distributed systems. Much like base CloudSim, DSLab IaaS's focus is on traditional machine virtualization. The simulator's standout feature is its ability to model multiple concurrently operating VM schedulers within the same host. The same authors also propose DSLab FaaS~\cite{semenov2024dslabFaas} for the simulation of the Function-as-a-Service paradigm. However, there is no evidence that the two simulators can be used together, even though they originate from the same underlying framework.
The primary drawback of both works is that they are written in Rust, a programming language with a much smaller community of developers than Java or C, due to the steep learning curve. On the contrary, much of CloudSim's popularity is to be attributed to the ease of use, whose only requirement is basic knowledge of the Java programming language.

FaaS-Sim~\cite{raith2023faasSim} is a Python-based, trace-driven stochastic simulation framework designed to evaluate placement and scaling decisions in serverless computing platforms. Given a network topology, a software architecture spanning the Edge-Cloud continuum, and workload traces, FaaS-Sim provides estimates of function execution times and resource usage. Compared to CloudSimSC, it provides a more accurate resource model validated with real-world traces in comparison to NS2.

OpenDC~\cite{mastenbroek2021opendc} is a datacenter simulation platform composed by: i) a web-based user interface to interactively construct, share, and reuse datacenter designs; ii) a collection of convenience tools to configure and automate experiments, and set up metric collectors and workload traces, among others; and iii) a model-driven discrete-event simulator written in Kotlin. Moreover, OpenDC comes with a collection of prefabricated scenarios, such as serverless and HPC infrastructures.  However, the platform specifically focuses on datacenter modeling, and it appears to be primarily designed for exploring prebuilt use cases in an educational context. Therefore, it is unclear how easily third-party researchers can extend it to meet their specific needs (i.e. simulating the Cloud-Edge continuum).

In conclusion, CloudSim has established itself as the tool of choice for researchers in this landscape thanks to: i) its ease of use, by requiring only beginner-level Java knowledge and minimal expertise in specific domains, such as networking or serverless computing; ii) its abstraction level, which is well-suited for simulating key characteristics of Cloud scenarios without having to perform a fine-grained packet-level, hop-by-hop, simulation in a distributed worldwide network, as it would be needed with approaches based on network simulators; and iii) its wide array of extensions covering any Cloud-related computing paradigm.

\begin{figure*}
    \centering
    \includegraphics[width=0.9\linewidth]{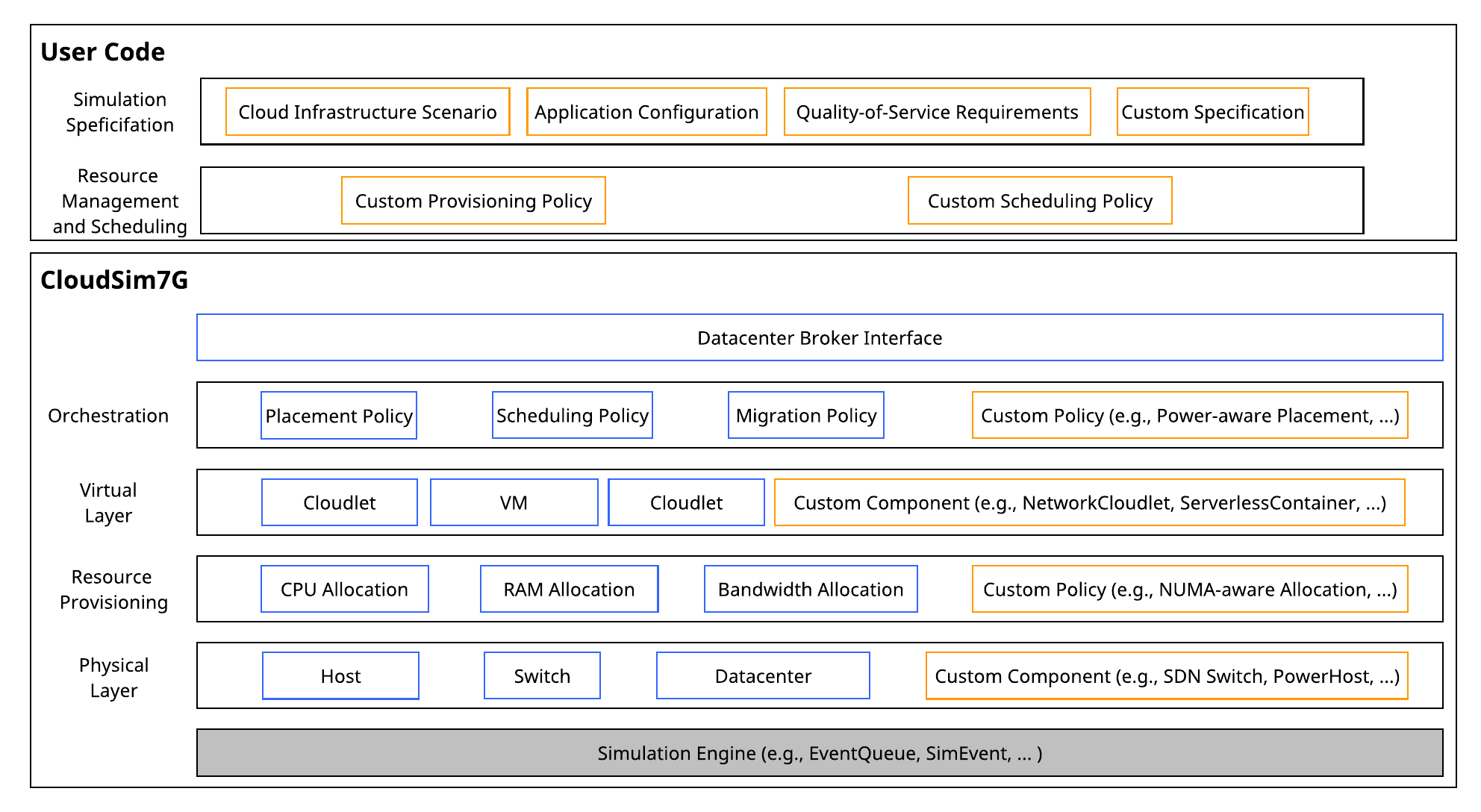}
    \caption{The architecture of CloudSim 7G. Blue boxes correspond to the CloudSim components of the base layer. Orange boxes correspond to user-defined policies and configurations, as well as newly created or extended CloudSim components.}
    \label{fig:cloudsim-arch}
\end{figure*}

\section{CloudSim 7G}\label{sec:cloudsim7g}
In CloudSim 7G, the codebase has undergone a massive refactoring and refinement process to accommodate the generalized interfaces that enable the coexistence of multiple modules within the same simulated scenario.
As a positive side-effect, the refactoring and refinement process allowed us to modernize and optimize the original codebase, improving the simulation performance, as well as its readability, usability, and flexibility. See~\Cref{sec:performance_eval} for a performance evaluation between CloudSim 6G and CloudSim 7G. Feature-wise, CloudSim 7G contains refined versions of many spin-off contributions previously bundled with CloudSim as independent packages. These contributions include containers, geo-located services, web load balancing, and network flow modeling. 

\Cref{subsec:arch} gives an overview of the high-level architecture of CloudSim 7G. Subsequently, \Cref{subsec:cloudsim-workflow} describes the process of running a simulation in practice, thus providing context for readers unfamiliar with CloudSim before delving into the technical changes introduced by CloudSim7G.
\Cref{subsec:design} explains in detail the new internal design of CloudSim to implement the integration layer within the building blocks of CloudSim. \Cref{subsec:refactoring} and \Cref{subsec:refinement} provide a practical description of the refactoring, optimization and refinement process the codebase has undergone to accommodate the adaptation interface. Lastly, \Cref{subsec:restore_compatibility} gives the guidelines to update an old module to CloudSim 7G.

\subsection{Architecture}\label{subsec:arch}
\Cref{fig:cloudsim-arch} shows the up-to-date architectural components of CloudSim 7G. Each building block has been refactored, and modernized, and several have been removed or consolidated (compared to the figure in the seminal paper~\cite{calheiros2011cloudsim}) thanks to the generalized interfaces of CloudSim 7G. The overall structure is the same: a user-facing section for Cloud providers, application developers, and researchers, called the ``User Code'', and a ``backend'' section dedicated to the development of the simulator and its modules. 

The bottom most layer is the simulation engine, which provides the core functionalities to start, pause, and stop simulated entities, as well as store and dispatch the discrete events to be processed at run-time. The subsequent layers consist of the building blocks required to characterize a Cloud infrastructure: from the physical layer (i.e., hosts, switches, etc.) up to the orchestration layer (i.e., VM placement strategies, cloudlet scheduling policies, etc.).    

The top most layer is the ``User Code'', which exposes the functionalities to customize each layer and evaluate a workload on the simulated infrastructure. More specifically, a user specifies: i) the characteristics of the Cloud infrastructure in terms of the number of physical hosts, their hardware specification, and how they are interconnected (if the user is interested in modeling the network flow); ii) the characteristics of the virtual components and their resource requirements; iii) the characteristics of the Cloud applications coupled with optional Quality-of-Service requirements; and iv) the characteristics of any installed CloudSim module. The latter are typically loaded into the user installation of CloudSim using build automation tools like Maven or Gradle. Depending on the modules used by the CloudSim user, an application may be expressed as a workflow (using NetworkCloudSim, for instance), or a host may exhibit energy-awareness capabilities through the power-aware module. A Cloud researcher may go a step further and programmatically extend such functionalities to perform a customized and more complex evaluation of resource management techniques. These are typically more ``aware'' provisioning, placement, or scheduling policies: a Cloud researcher may implement its own topology-aware cloudlet scheduling policy to optimize a workflow application within a networked datacenter, or exploit the migration module to develop an auto-scaling policy for web sessions. 

Subsequently, CloudSim configures the simulation according to the user specifications, replacing the basic building blocks with the custom policies and components expressed in the user code. In particular, CloudSim 7G provides out-of-the-box support for containers, VMs, ``traditional'' cloudlets, and workflow applications (i.e., networked cloudlets constructed as graphs) at the virtual layer. The latter resources are then orchestrated through a collection of basic placement, scheduling, and migration policies. For instance, CloudSim's power-aware package uses migration policies to optimize the VM placement using statistical analysis of host load utilization. Throughout the software stack, it is possible to exploit custom components, such as the serverless capabilities of CloudSimSC, as long as they are compatible with CloudSim 7G. 

\subsection{Running a Simulation in CloudSim}\label{subsec:cloudsim-workflow}
A simple scenario simulated with base CloudSim consists of the following steps: the service broker submits an inventory to the datacenter, which comprises the virtual machines (VMs), physical hosts, and a list of ephemeral activities to be performed. The latter abstracts the concept of a request submitted to a Cloud service and in CloudSim jargon they are called cloudlets.
Then, the datacenter deploys the submitted inventory using a variety of resource provisioning methods and scheduling policies, both at the VM and cloudlet levels. The simulation terminates when all the submitted activities have been executed. The physical hosts and VMs are specified in terms of the number of processing elements, available RAM, and network bandwidth. Each processing element (PE) has a specific processing strength measured in millions of instructions per second (MIPS), and the capabilities of the physical host limit the hardware resources allocated to each VM.

A cloudlet specifies an execution length, in terms of millions of instructions (MI), and the number of required processing elements to be provisioned.  The actual execution time is determined by the capabilities of the underlying VM, the number of co-hosted activities on it, and the cloudlet scheduling policy. Out-of-the-box CloudSim supports two scheduling policies: space-shared, where only one cloudlet at a time can execute (others will be on a waiting list), and time-shared, where the processing strength is shared among cloudlets running simultaneously. More specifically, for time-shared scheduling, the start time of a cloudlet corresponds to the submission time, since there is no queuing, and the estimated finish time solely depends on the current processing capacity. For space-shared scheduling, the estimated start time depends on the cloudlet's position in the waiting list, whereas the current processing capacity of the VM is constant since always only one cloudlet at times is executing. Refer to the seminal paper at~\cite{calheiros2011cloudsim} for a more in-depth description of the simulated entities and their interactions.  
The power of CloudSim comes from the framework's extensibility, which allows researchers to customize existing components and policies, or create new ones, at different levels of the infrastructure (see \Cref{fig:cloudsim-arch}).

\subsection{Design}\label{subsec:design}
\begin{figure*}
\centering
\begin{subfigure}{\textwidth}
      \centering
      \includegraphics[width=\linewidth]{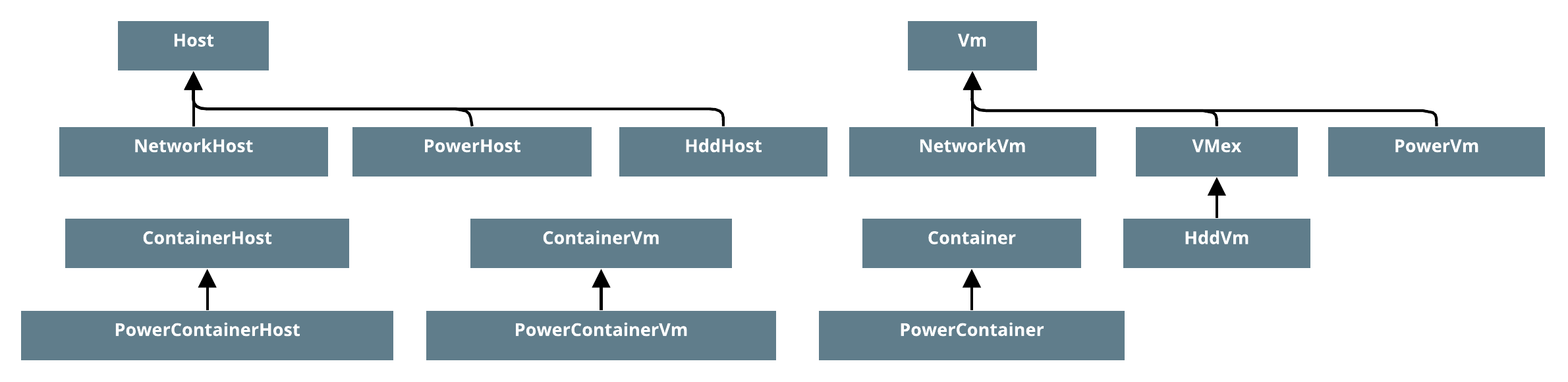}
      \caption{CloudSim 6G}
      \label{subfig:cloudsim6g-entity-uml}
\end{subfigure}
\hfill
\begin{subfigure}{\textwidth}
  \centering
  \includegraphics[width=\linewidth]{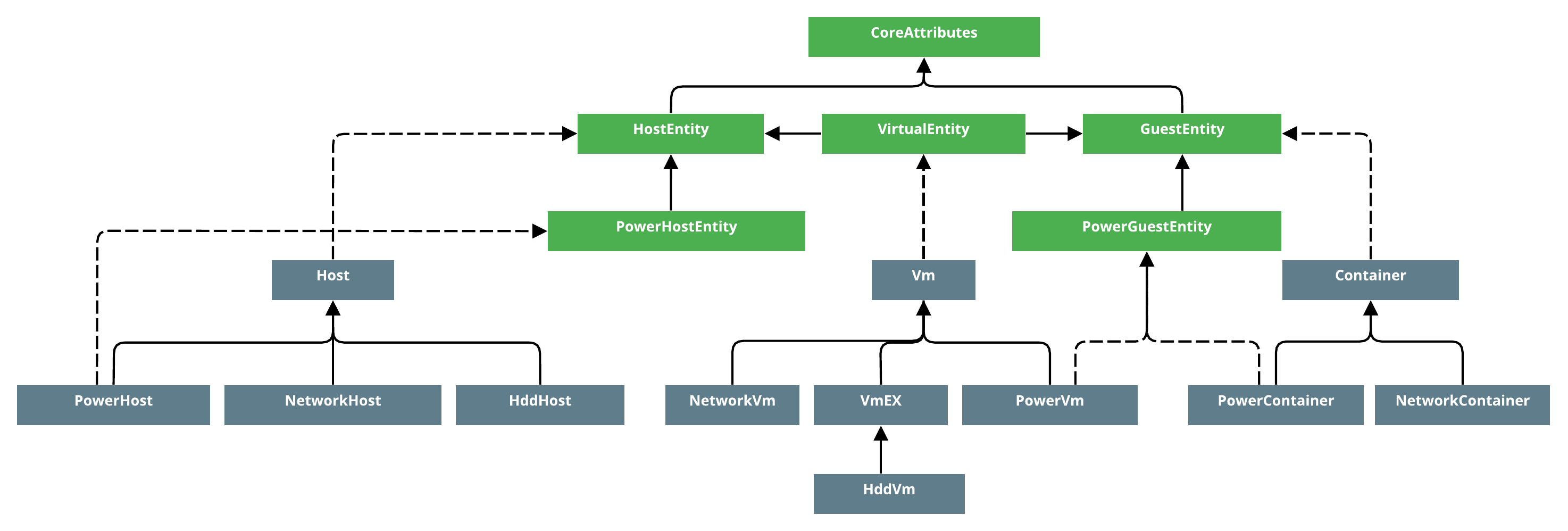}
  \caption{CloudSim 7G}
  \label{subfig:cloudsim7g-entity-uml}
\end{subfigure}
\caption{UML class diagram of host and guest entities in CloudSim. Green boxes depict Java interfaces, and blue boxes represent Java classes. The relationship arrows follow the UML relations notation style: a dashed line represents a realization/implementation, and a full line represents inheritance.}
\label{fig:cloudsim-entity-uml}
\end{figure*}
CloudSim is a toolkit that leverages the features of the Java programming language to offer a highly extensible and portable simulation platform. CloudSim modules are developed using inheritance and composition in Java to enhance existing components or create new ones.
The design changes of CloudSim 7G affected the following Java classes (and their extensions): \texttt{Datacenter}, \texttt{DatacenterBroker}, \texttt{Host}, \texttt{Vm}, \texttt{Container}, \texttt{VmAllocationPolicy}, and \texttt{CloudletScheduler}. These classes implement the building blocks of the simulator depicted in \Cref{fig:cloudsim-arch}.

CloudSim 7G presents a collection of Java interfaces that generalize the definition of some components and facilitate the coexistence of multiple CloudSim modules within the same simulated scenario. This design change is one of the biggest shifts in the codebase. Firstly, CloudSim 7G introduces the concept of \textit{Guest} entity and \textit{Host} entity. A guest entity executes actions related to the management and processing of cloudlets according to a given scheduling policy. A host entity executes actions related to the management (i.e., allocation, provisioning, and scheduling) of guest entities within the simulated Cloud infrastructure. A guest entity runs inside a host entity which may be shared among several guest entities. Such sharing affects the ability of a guest entity to execute its actions. \Cref{fig:cloudsim-entity-uml} depicts the UML class diagram of the components affected by the design change.
For instance, the authors of ContainerCloudSim had to distinguish between the \texttt{Container} and \texttt{Vm} classes. The two components performed the same actions but were implemented as completely independent entities. This triggered the need for further redundant components: an instance of the \texttt{Container} class required an extension of the \texttt{Vm} class, called \texttt{ContainerVm} (i.e., a VM that hosts containers). In turn, the latter required an extension of the \texttt{Host} class called \texttt{ContainerHost}, which can only reside on an extension of the \texttt{Datacenter} class, the \texttt{ContainerDatacenter}. 
To overcome such structural limitations of the framework, CloudSim 7G introduces the following Java interfaces:
\begin{enumerate}
    \item \texttt{HostEntity}: A Java interface featuring a collection of abstract methods required to implement a host entity and its actions. A few methods provide a default, generalized implementation of an action. Therefore, a possible implementation of the \texttt{HostEntity} interface corresponds to the \texttt{Host} class in previous CloudSim versions.  
    \item \texttt{GuestEntity}: A Java interface featuring a collection of abstract methods required to implement a guest entity. Similarly, a few generalized implementations are provided by default. Therefore, a possible implementation of the \texttt{GuestEntity} interface corresponds to the \texttt{Vm} and \texttt{Container} classes in previous CloudSim versions.
    \item \texttt{CoreAttributes}: A Java interface that defines the methods required for implementation by both \texttt{HostEntity} and \texttt{GuestEntity} classes.
    \item \texttt{VirtualEntity}: A placeholder interface to implement an entity that is simultaneously a \texttt{HostEntity} and a \texttt{GuestEntity}. This interface is essential to support nested virtualization.

    \item \texttt{PowerHostEntity} and \texttt{PowerGuestEntity}: Extended interfaces to standardize the integration of power-aware features into the implementation of host and guest entities.
\end{enumerate}
The new design has made certain components entirely redundant, such as the \texttt{ContainerHost} and \texttt{ContainerVm} classes (see~\Cref{fig:cloudsim-entity-uml}), while repurposing others. For example, the \texttt{NetworkVm} class, which served no functional purpose, has been redesigned to incorporate the new virtualization overhead feature, which is discussed in \Cref{subsec:refinement}. As a result, module developers must now adhere to the newly defined Java interfaces to be compatible with CloudSim 7G and leverage its multi-extension support (More on this in \Cref{subsec:restore_compatibility}).

\begin{figure*}
\centering
\begin{subfigure}{\textwidth}
      \centering
      \includegraphics[width=\linewidth]{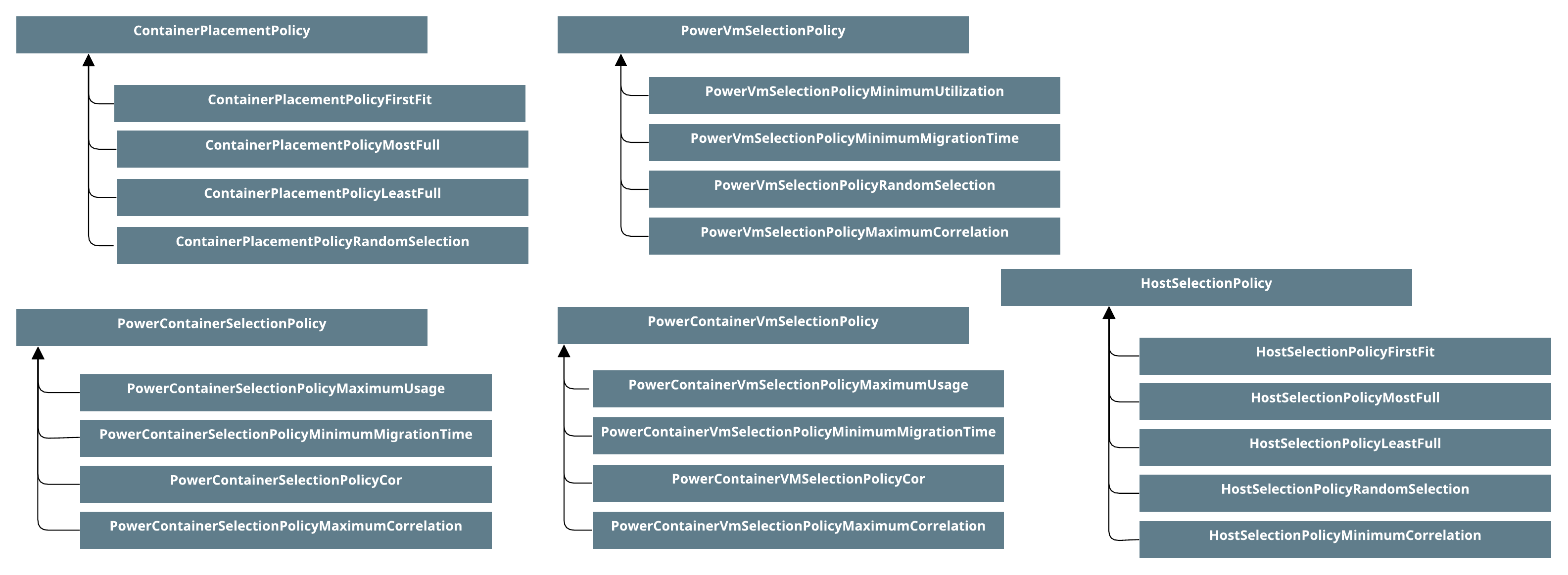}
      \caption{CloudSim 6G}
      \label{subfig:cloudsim6g-selection-uml}
\end{subfigure}
\hfill
\begin{subfigure}{.8\textwidth}
  \centering
  \includegraphics[width=\linewidth]{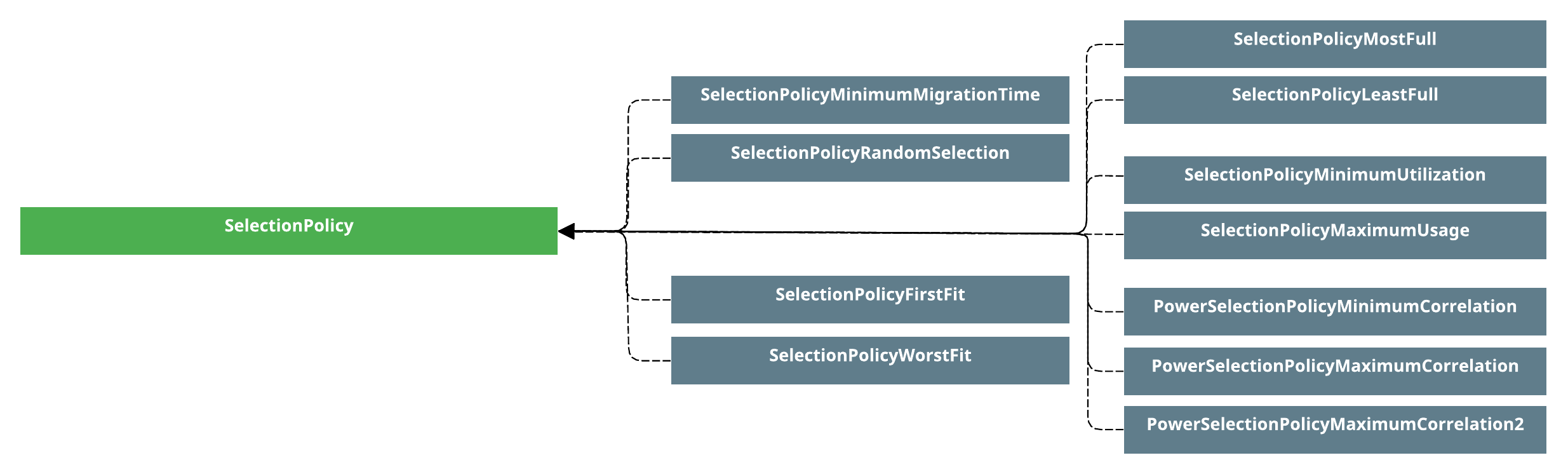}
  \caption{CloudSim 7G}
  \label{subfig:cloudsim7g-selection-uml}
\end{subfigure}
\caption{UML class diagram of the host and guest selection policies in CloudSim. Green boxes depict Java interfaces, and blue boxes represent Java classes. The relationship arrows follow the UML relations notation style: a dashed line represents a realization/implementation, and a full line represents inheritance. Line colors are for presentation purposes only.}

\label{fig:cloudsim-selection-uml}
\end{figure*}

Secondly, CloudSim 7G introduces the concept of \textit{selection policy}, which generalizes the process of selecting an entity with a criterion from within a list of candidates. This concept is typically employed to implement placement and migration policies. In previous iterations of CloudSim,  there is a clear divergence between a placement policy (i.e., select a host for a VM to be placed) and a migration policy (i.e., select a VM to be migrated), despite being essentially the same activity. Therefore, the selection policy interface in CloudSim 7G serves as the fundamental building block for implementing placement, migration, or any policy that involves selecting an entity. This design change significantly simplifies the codebase, as illustrated in \Cref{fig:cloudsim-selection-uml}, by rendering redundant several Java classes from ContainerCloudSim and the power module. Specifically, the amount of Java classes related to the selection process decreases from 26 to 11.

In conclusion, CloudSim 6G features a large number of unnecessary copied-pasted classes. As previously stated, the original version of ContainerCloudSim used to treat VMs and containers as completely independent entities, even if they performed the same actions. In comparison, CloudSim 7G provides a common collection of interfaces that unites most modules developed so far, as long as they are developed correctly using OO principles. The new internal design of CloudSim 7G helped with code deduplication: thanks to the new architecture, ContainerCloudSim has undergone a 64\% reduction in terms of line of code (LoC). Similarly, the source code of NetworkCloudSim has been reduced by 50\%, and the power module has been reduced by 21\%. 
The next two sections provide a detailed explanation of the significant changes to the codebase, in terms of code deduplication, improving time and space complexity, and adapting the existing components to the new internal design.

\subsection{Code Refactoring and Optimization}\label{subsec:refactoring}
The CloudSim refactoring process focused mainly on code modernization and deduplication: 
\begin{enumerate}
    \item Removed multiple parts of the codebase that contained redundant operations, such as leftover code from GridSim~\cite{buyya2002gridsim} and were marked as deprecated.

    \item Removed the synchronized keyword from some data structures, given the single-threaded nature of the core simulation engine.

    \item Removed duplicated and redundant code snippets within extensions to CloudSim's core components by leveraging the new internal design described in the previous section. This deduplication process encompassed most components, such as \texttt{Datacenter}, \texttt{Vm}, and \texttt{DatacenterBroker}, as well as the scheduling and selection components.

    \item Performed the necessary code upgrades for supporting the latest Java Development Kit (JDK), the JDK21, as well as JUnit 5.
\end{enumerate}
The total amount of LoC removed is more than 13000, with no functionality lost during the deduplication process. This was verified by running tests provided in the CloudSim example folder. Moreover, these code changes improve code reusability and favor the enforcement of compliance with OO principles.

The CloudSim code optimization process focused on selecting the proper data structures and coding practices to improve memory usage and reduce run-time, without altering the basic CloudSim's functionalities. Indeed, depending on the use-case, choosing the appropriate data structure significantly improves the performance of the simulator. For example \texttt{ArrayList} and \texttt{LinkedList} have strengths and weaknesses in different operations: \texttt{Iterator.remove()} and \texttt{add()} have constant complexity in a \texttt{LinkedList} but linear complexity in an \texttt{ArrayList}; on the other hand, the \texttt{get()} operation has linear complexity in a \texttt{LinkedList} but constant one in an \texttt{ArrayList}, on average. 

More specifically, CloudSim 7G introduces the following code optimizations:
\begin{enumerate}    
    \item The simulation core engine previously used a custom linked-list for event dispatching to simulation entities, resulting in an inefficient $O(n)$ time complexity for maintaining temporal order. This implementation has been replaced with Java's \texttt{PriorityQueue}, which guarantees $O(logn)$ time complexity for queueing methods.
        
    \item Enforced the use of method \texttt{isEmpty()} for lists, instead of checking the ``actual'' size, as in certain implementations (such as \texttt{LinkedList}), determining the size requires counting the items, which is more time-consuming than simply checking for emptiness.
    
    \item Optimized string operations by using \texttt{StringBuffer} and \texttt{StringBuilder}. During a simulation, the simulator prints many logs on the console or outputs them to log files. For convenience, many developers use the plus operator to concatenate strings. However, since Java strings are immutable, this coding behavior caused unnecessary load on the CPU and memory.

    \item CloudSim previously used an \texttt{ArrayList} to store historical data about CPU utilization for power-awareness. However, since only the last item is frequently accessed, and new items are simply appended, a \texttt{LinkedList} is a more efficient choice. 
        
    \item Preferred the use of primitive data types (e.g., \texttt{int}, \texttt{long}, \texttt{double}) rather than classes (e.g., \texttt{Integer}, \texttt{Long}, \texttt{Double}) where possible to minimize object creation and auto-boxing/auto-unboxing overheads. 
    
    \item Maximized the reuse of objects to avoid the frequent creation/destruction of objects from the Java heap. This optimizes memory usage and reduces the frequency and load of garbage collection (GC).

    \item Frequently invoked methods with non-constant computational complexity were cached in local variables for efficiency. For instance, the calculation of the required per-core MIPS of VMs and containers used to be performed continuously throughout the simulation, which involved iterating through multiple arrays. Similarly, \texttt{getUid()} previously re-constructed the unique identifier from scratch with each call, involving a string concatenation.
\end{enumerate}
Refer to~\Cref{sec:performance_eval} for a performance evaluation between CloudSim 6G and CloudSim 7G, showcasing the improvements.
 
\subsection{Code Refinement}\label{subsec:refinement}
The code refinement process involves more than just simple code deduplication: it requires proper adaptation of the codebase to fully leverage the new internal design, in the attempt to enhance both user and developer experience when using CloudSim. In particular, we targeted: i) the event handling system within the simulation core engine; ii) the main logic of the \texttt{CloudletScheduler} class; and iii) the entire NetworkCloudSim module. Detailed descriptions of each bullet point are provided in the following paragraphs, in order. 

CloudSim employs a set of event tags to indicate an action that needs to be undertaken by a simulated entity when an event is triggered at run-time. Each CloudSim module typically creates its own collection of custom tags, which can potentially lead to collisions if integer values are inadvertently reused, to the detriment of multi-module scenarios. In CloudSim 7G, the event handling system has been updated to make use of the \texttt{Enum} Java class for declaring tags to improve code readability and prevent collisions.

The original version of \texttt{CloudletScheduler} abstract class was not a generic template that represented the life-cycle of a cloudlet deployed on a guest entity: any extension to the \texttt{CloudletScheduler} class needed to re-implement the whole scheduling logic, causing a lot of redundant code. As a significant side-effect, different implementations of the \texttt{Cloudlet} class could not coexist within the same cloudlet scheduler, and consequently, on the same VM (unless manually managed by the module developer, for each possible cloudlet implementation). CloudSim 7G solves this problem by providing a revised cloudlet scheduling life-cycle represented by \Cref{alg:cloudletsched_lifecycle}, extracted from the \texttt{CloudletScheduler} abstract class. Lines 4, 7, and 14 expose 3 handler methods (highlighted in gray) to standardize the customization of the scheduling behaviors. The first two handlers are used to customize the update logic and stopping condition of a cloudlet. Therefore, any extension to the \texttt{Cloudlet} class is supported out-of-the-box by a \texttt{CloudletScheduler} instance. For instance, the \texttt{NetworkCloudlet} class exploits these 2 handlers to implement the stages that realize an activity within a scientific workflow. The cloudlet scheduler uses the third handler to customize the unpause logic for cloudlets that are waiting to be executed.   
Lines 1-6 update the number of executed instructions of each active cloudlet in the time interval since the previous processing update. Lines 7-10 terminate the cloudlets that finished executing all the instructions. Line 11-13 returns early if the cloudlet scheduler has no more events (i.e., no active cloudlets) in the foreseeable future. Lines 14-16 unpause a subset of waiting cloudlets based on a customizable behavior of the cloudlet scheduler. For instance, \texttt{CloudletSchedulerTimeShared} class does not use the handler method, since all the submitted cloudlets are executed in a time-shared manner. Lines 17-23 estimate the simulation time for the next discrete event for this cloudlet scheduler, which depends on the finish time of the earliest finishing cloudlet. 
Thanks to this refinement process, the \texttt{CloudletScheduler} class and its extensions have undergone a 40\% LoC reduction.

\begin{algorithm}
	\caption{Revised logic of the cloudlet scheduler instantiated by the guest entities}\label{alg:cloudletsched_lifecycle}
	\begin{algorithmic}[1]
            \Require{$mipsShare$, currently available processing power to the guest entity}
		\Require{$currentTime$, current simulation time}
            \Require{$previousTime$, simulation time of the previous scheduling update}
            \Require{$cloudletExecList$, list of active cloudlets on the guest entity}
            \Require{$cloudletWaitList$, list of waiting cloudlets on the guest entity}
            \Ensure{Predicted completion time of the earliest finishing cloudlet, or $0$ if there are no active cloudlets left}
		\State{$timeSpan \gets currentTime - previousTime$}
		\For{$cl \in cloudletExecList$}
                \State{$allocMips \gets currentlyAllocatedMipsForCloudlet(currentTime, mipsShare, cl)$}
			\State{$\colorbox{lightgray}{cl.update()}$}
				\State{$cl.cloudletLengthSoFar \gets cl.cloudletLengthSoFar + (timeSpan \cdot allocMips)$}

                \If{$\colorbox{lightgray}{cl.isFinished()}$}
				\State{$cloudletExecList.remove(cl)$}
			\EndIf
		\EndFor
  
            \If{$cloudletExecList.empty() \wedge cloudletWaitList.empty()$ }
			\State\Return{0}
		\EndIf
  
            \State{$unpausedCloudletList \gets \colorbox{lightgray}{unpauseCloudlets(cloudletWaitList)}$}
            \State{$cloudletWaitList.removeAll(unpausedCloudletList)$}
            \State{$cloudletExecList.addAll(unpausedCloudletList)$}

            \State{$nextEvent \gets Double.MAX\_VALUE$}
            \For{$cl \in cloudletExecList$}
                \State{$estFinishTime \gets currentTime + \frac{cl.cloudletLength - cl.cloudletLengthSoFar}{currentlyAllocatedMipsForCloudlet(currentTime, mipsShare, cl)}$}
                \If{$estFinishTime < nextEvent$}
                    \State{$nextEvent \gets estFinishTime$}
                \EndIf
		\EndFor
            \State\Return{$nextEvent$}
	\end{algorithmic}
\end{algorithm}

Regarding NetworkCloudSim, it introduced a series of network-related features that were cumbersome to set up, and some features were not fully implemented, leading to confusion and difficulty in their use. Moreover, there were plenty of hard-coded constants that could not be customized. In short, NetworkCloudSim is presented as a proof of concept, rather than a usable module. To this end, CloudSim 7G includes a revised version of NetworkCloudSim, which has been generalized to function as a proper module, and enriched with several new features to complete and enhance the user experience, such as the introduction of virtualization overhead.

For instance, there was no user-friendly way to configure a network with multiple switches. As a result, a CloudSim user had to directly access the member variables of a \texttt{Switch} instance, which was very error-prone. Furthermore, the \texttt{NetworkCloudlet} class, which implemented the cloudlets with network capabilities for constructing workflow applications, followed a different execution model than the traditional cloudlet, despite inheriting the properties from the \texttt{Cloudlet} class. A stage within a networked cloudlet was defined in terms of execution time (in milliseconds), whereas a traditional cloudlet is defined in terms of execution length (in millions of instructions). The payload size of a packet was defined in bytes, but it was not converted to bits when calculating the packet transmission time. NetworkCloudSim allowed configuring a deadline to the execution time of the workflow application, but it did not implement the logic to check whether the deadline had been met.

Finally, the virtualization overhead feature introduces an optional delay applied each time a packet traverses a guest or host entity. This feature is especially useful for assessing resource management techniques across different virtualization deployments, including the nested virtualization capabilities introduced with CloudSim 7G, as demonstrated in \Cref{tab:use_case}. Without accounting for virtualization overhead, CloudSim would fail to accurately capture the performance differences between virtual machines (VMs), containers, and nested virtualization.

\subsection{Guidelines for Upgrading a Module to CloudSim 7G}\label{subsec:restore_compatibility}
This section provides the guidelines for restoring the compatibility of older CloudSim modules with the new base layer of CloudSim 7G. Most re-engineered Java methods within classes for core CloudSim components have retained their original APIs, some kept the previous one as a deprecated alternative.
While we aimed to minimize disruptions, the new internal design of CloudSim 7G necessitated key changes that need to be addressed by extension developers; Using modern IDEs should make the upgrade process piloted and straight-forward, as they can detect dangling code paths and the use of deprecated methods.

Firstly, some classes have been removed or replaced: for instance, the \texttt{ResCloudlet} class has been completely integrated into the \texttt{Cloudlet} class. Consequently, any extension to the \texttt{ResCloudlet} class now needs to inherit its properties from the \texttt{Cloudlet} class. Secondly, some methods using Java generics require conversion from type \texttt{Host} to \texttt{HostEntity} (or from \texttt{Vm} to \texttt{GuestEntity}) due to ambiguities caused by the type erasure mechanism.

Extensions to allocation and migration policies must adhere to the novel API offered by the selection policy interface.
Lastly, since the tag system in CloudSim has been entirely rewritten to use the \texttt{Enum} class instead of integer primitives, extension developers will need to update their custom tags accordingly.
As a side note: although the novel template-based \texttt{CloudletScheduler} class maintains full backward compatibility, we encourage module developers to adhere to the new framework outlined in \Cref{alg:cloudletsched_lifecycle}, thus integrating their custom changes to the scheduling life-cycle through the proposed handler methods.
\section{Performance Evaluation}\label{sec:performance_eval}

\begin{table}[]
    \centering
    \begin{tabular}{|c||c|c|c||c|c|c|}
    \hline
    Algorithm
            & \multicolumn{3}{c||}{Total Heap Usage (MB)} & \multicolumn{3}{c|}{Run-time (s)}\\    &   6G  &   7G & \textbf{Improvement}  &   6G  &   7G  & \textbf{Improvement} \\
    \hline\hline
    Dvfs 
    &   2590.895  &  2092.710 &  19\% &  8.863 & 7.787 & 12\%\\
    \hline
    MadMmt 
    &   43139.817  & 42299.257 &  2\% &   88.666 & 84.333 & 5\% \\
    \hline
    ThrMu  
    &   42404.056 & 33157.480 &  22\% &   83 & 73 & 12\% \\
    \hline
    IqrRs  
    &   41237.034  & 31038.492 &  25\% &  84.666  &  74.333 & 12\%\\
    \hline
    LrrMc 
    &   40777.280  & 33584.612  &  18\% &  79.333 & 70.666 & 11\%\\
    \hline
\end{tabular}
    \caption{Performance comparison of CloudSim 6G and CloudSim 7G in terms of total allocated memory and run-time. Each experiment is performed 3 times, and the average result is shown in the table. The ``Improvement'' column displays the percentage decrease for each metric.}
    \label{tab:perf_eval}
\end{table}

In CloudSim 7G, the codebase has undergone massive refactoring to improve the readability and performance of the code.
This section presents a series of experiments conducted on an x86-64 server with Intel(R) Xeon(R) Gold 6238R CPU, 80MB L3 cache, 58MB L2 cache, 128GB RAM, and Ubuntu 22.04 LTS. The goal is to evaluate the performance improvements of CloudSim 7G, compared to CloudSim 6G, using large trace datasets, despite the introduction of new abstraction layers. 
We utilized OpenJDK21 in both CloudSim 6G and CloudSim 7G, namely OpenJDK 64-Bit Server VM build 21.0.5+11-Ubuntu-1ubuntu122.04.
All experiments are conducted with JVM parameters \texttt{-Xms4G -Xmx4G} (i.e., fixed heap size to avoid dynamic resizing at run-time) using the default garbage-first garbage collector~\cite{detlefs2004g1gc}. 

We collected the total memory allocated by each experiment at run-time by aggregating the reclaimed heap memory reported in Java Garbage Collection log messages (i.e., JVM parameter \texttt{-Xlog:gc*}).
The wall-clock run-time is obtained from the execution summary provided by Maven, the build automation tool used by CloudSim.
All experiments are conducted with CPU frequency blocked at 2.20 GHz (the base frequency suggested by the vendor), and no turbo or dynamic boosting.
For each algorithm and metric combination, we conducted the experiment 3 times and computed the average heap usage and wall-clock run-time.

For the sake of reproducibility, we evaluate the performance improvements of CloudSim 7G using a collection of simple heuristic algorithms from the power module\cite{beloglazov2012optimal} and workload traces sampled from the PlanetLab package~\cite{park2006comon}, both of which are readily available within the CloudSim package. Each experiment involves a distinct VM allocation and selection policy for the study of energy and performance-efficient dynamic consolidation of VMs within a Cloud datacenter.

\Cref{tab:perf_eval} depicts the results of the performance comparison: CloudSim 7G consistently outperforms CloudSim 6G in terms of average total heap usage and wall-clock run-time. In selected experiments, CloudSim 7G reduces run-time by up to 9 seconds and lowers the total memory allocated by as much as 10000 MB, greatly improving the performance of the simulator.

\section{Case Study}\label{sec:case-study}

This section presents a simulation scenario that uses multiple CloudSim modules. In particular, the following entities will interact in a multi-module scenario: i) ``traditional'' VMs provided by CloudSim since the first version; ii) containers from the refactored version of ContainerCloudSim; iii) the network, virtualization overhead, and workflow application model of the overhauled NetworkCloudSim, and iii) the service broker of CloudSimEX for simulating realistic cloudlet arrivals. We will use the term ``guest'' to refer to a VM or container as a general virtual component, ``nested'' when a container is placed on a VM instead of a physical host directly, and denote the individual virtualization technologies by the symbols $V$, $C$ and $N$, respectively, when necessary.
Notice that the goal of this section is not to thoroughly evaluate a set of placement or scheduling strategies to optimize a deployment,
but to demonstrate CloudSim 7G's new multi-module feature with a simple scenario.

In the following, we take the viewpoint of a private Cloud provider that needs to deploy a workflow application of interconnected tasks, represented as a simple Directed Acyclic Graph (DAG). The provider's goal is to estimate the end-to-end (E2E) deadline for its application, formally called the $makespan$ of the DAG, based on some known parameters: i) application-wise, the number of instructions per task, the payload to be transferred between tasks, and the inter-arrival time of new requests; and ii) infrastructure-wise, the datacenter topology, the link bandwidths, and the overhead introduced by virtualization when using the network. 
This use-case is notoriously faced by telco providers in the context of Industry 4.0, where time-critical applications, characterized by strict timing and reliability constraints, are being deployed on contemporary Cloud infrastructures. The multi-tenant nature of traditional Cloud infrastructures poses a challenge for such an emerging use-case due to the ``noisy neighbor'' problem~\cite{abeni2023rtCloud, garcia2014challengesInRtCloud}. Notice that previous versions of CloudSim do not support the simulation of such a peculiar scenario. 

In our experimental evaluation, we will analyze how the makespan is affected by varying the cloudlet placement strategy, the network delay, the payload size, the overhead introduced by different virtualization configurations (i.e., VMs, containers, or nested containers), and the computational noise of co-located cloudlets.

\begin{figure*}
\centering
\begin{subfigure}{.3\textwidth}
      \centering
      \includegraphics[width=0.8\linewidth]{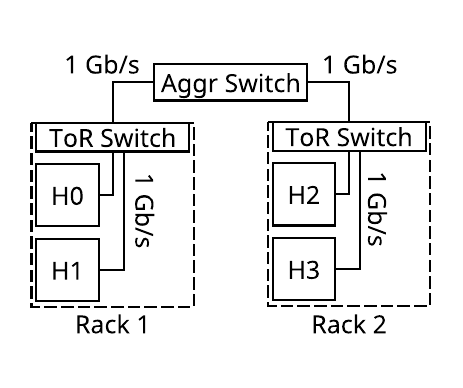}
      \caption{Simulated datacenter.}
      \label{subfig:dc}
\end{subfigure}%
\hfill
\begin{subfigure}{.3\textwidth}
      \centering
      \includegraphics[width=0.8\linewidth]{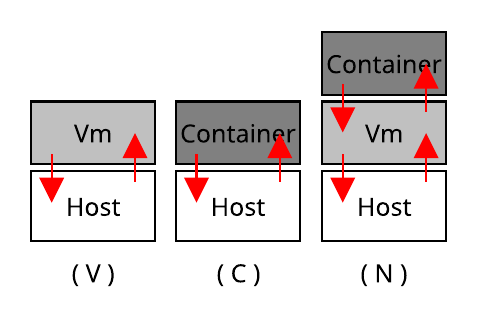}
      \caption{Virtualization configurations.}
      \label{subfig:virt}
\end{subfigure}%
\hfill
\begin{subfigure}{.3\textwidth}
  \centering
  \includegraphics[width=0.8\linewidth]{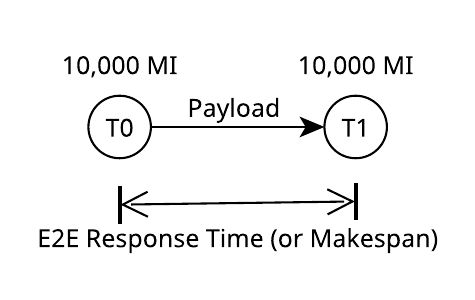}
  \caption{Simulated workflow application.}
  \label{subfig:dag}
\end{subfigure}
\caption{Case-study: A simple workflow application with end-to-end deadline deployed on a datacenter of interconnected physical hosts.}
\label{fig:cloudsim-case-study}
\end{figure*}
The private datacenter in \Cref{subfig:dc} comprises 4 homogeneous physical hosts interconnected via 2 switches. The network topology resembles a basic tree-like structure: the hosts are split equally between 2 different racks, and each group of hosts is connected to a Top-of-Rack (ToR) switch using a symmetrical gigabit connection. The ToR switches communicate through an Aggregate Switch via a symmetrical gigabit connection as well. 
\Cref{subfig:virt} showcases the type of virtualization deployment under study, each distinguished by a corresponding symbol, with $N$ denoting the ``Container-on-VM'' nested virtualization.

The logical representation of the workflow application is depicted in \Cref{subfig:dag}: a simple DAG with two tasks, $T0$ and $T1$, the source and sink node of the graph, respectively,  connected in a chain by a data transfer dependency. The DAG is ``activated'' by user requests that trigger periodically. Therefore a DAG activation is a periodic event that refers to the following chain of activities: i) $T0$ and $T1$ are submitted for execution to the datacenter; ii) $T0$ executes for a pre-fixed amount of millions of instructions $L_{T0}$; iii) $T0$ transmits a payload to $T1$ of fixed-size $payloadSize$; and iv) $T1$ executes for a pre-fixed amount of $L_{T1}$. The requests are simulated in CloudSim by submitting two networked cloudlets interconnected according to the application topology.

For the sake of simplicity, we will only focus on the effect of cloudlet scheduling, placement, and virtualization overhead (as depicted in \Cref{subfig:virt}), using two payload sizes for data transmission. In particular, the experimental evaluation focuses on 3 possible placement configurations for the cloudlets: I) $T0$ and $T1$ are co-located, meaning that data is transmitted locally without using the physical network; II) $T0$ and $T1$ are located on the same rack, but different physical hosts, meaning that data is routed through a ToR switch only; and III) $T0$ and $T1$ are located on different racks so that data must be transmitted through the aggregate switch as well. 

We assume physical hosts with hardware capabilities that meet the guests' requirements in terms of CPU, memory, and network bandwidth. Additionally, the guests are big enough to accommodate together all the cloudlets representing the workflow application, if necessary. \Cref{tab:use_case} summarizes the parameter configurations used for the experiment evaluation. We employ time-shared cloudlet scheduling to simulate computational noise for co-located cloudlets. The inter-arrival time of the cloudlets is sampled from an exponential probability distribution with rate parameter $\frac{L_{T0}}{mips_V} + \frac{L_{T1}}{mips_V} = 2.564$ so that there are overlaps between DAG activations only when the network is in use.

\begin{table}[]
    \centering
    \begin{tabular}{|c||c|c|c|}
    \hline
    & VM & Container & Nested Virt. \\\hline\hline
    Alloc. Processing Power & \multicolumn{3}{c|}{$mips_{V}=mips_{C}=mips_{N}=7800$ MIPS} \\\hline
    Alloc. Network Bandwidth & \multicolumn{3}{c|}{$bw_V = bw_C = bw_N = 1$ Gb/s} \\\hline
    Virt. Overhead& $O_V=5$ s & $O_C=3$ s & $O_N = O_V + O_C$ \\\hline
    Cloudlet Scheduler & \multicolumn{3}{c|}{Time-Shared} \\\hline
    Guest Scheduler & Time-Shared & - & Time-Shared \\
    \hline
\end{tabular}%
\begin{tabular}{|c||c|}
    \hline
    & Application \\\hline\hline
    Topology & $T0 \rightarrow T1$ \\\hline
    Execution Length& $L_{T0} = L_{T1} = 10000$ MI\\\hline
    Negligible payload& 1 bytes\\\hline
    Non-Neglibile payload& 1 GB\\\hline
    Request Inter-Arrival& $Exp\,(2.564)$
    \\\hline
\end{tabular}
    \caption{Simulation parameters used in the experimental evaluation.}
    \label{tab:use_case}
\end{table}
The VMs are modeled after the new AWS ``m7g.medium'' general-purpose EC2 instances\footnote{See: \url{https://aws.amazon.com/ec2/instance-types/m7g/}.}, which are single-core VMs powered by the AWS Graviton3 processor where the virtual CPU is a physical core running at 2.6 GHz. Containers are modeled as a ``slice'' of the underlying hosting component, which can be either a VM or the physical host, depending on the configuration; However, for the sake of simplicity, such slice corresponds to the entire hosting VM's compute capability when in nested virtualization.
The processing powers can be approximately estimated in MIPS using the following formula, which is derived from the textbook definition of CPU time~\cite{patterson2016Organization}:
\begin{align}\label{eq:mips}
    mips_V &= mips_C = mips_N = \frac{clk\_rate \cdot IPC}{10^6} = 7800 \text{ MIPS, assuming $IPC=3$}
\end{align}
where $clk\_rate$ is the CPU frequency in Hz, and $IPC$ is the number of instructions per cycle.
The execution length of the cloudlets is configured as $L_{T0} = L_{T1} = 10 000$ MI to ensure a sufficiently long completion time for presentation purposes. For the same reason, the virtualization overhead is configured to be exceedingly high (i.e., over one second).

The deployed workflow application must meet the desired E2E latency specified by the Cloud provider as a deadline. 
The tightest possible deadline equals the makespan of a DAG activation, which takes into account the compute times, as well as the transmission times due to the network switching delay. It can be estimated as:
\begin{align}\label{eq:makespan}
    M_{\alpha} = \sum_{i}^{\mathcal{T}} \left(\frac{L_i}{mips_{\alpha}} + \rho \cdot O_{\alpha}\right) + networkHops \cdot\sum_{i}^{\mathcal{T}}  \left(\frac{payloadSize}{bw_{\alpha}}\right) \text{, where } \rho &=
        \left\{\begin{array}{ll}
                1 & networkHops > 0 \\[+0.1cm]
                0 & \text{otherwise}
        \end{array}\right.,
\end{align}
where $\mathcal{T} = \{T0, T1\}$ denotes the set of tasks in the DAG, and $\alpha \in \{V,C,N\}$ represents the virtualization configuration. The symbol $L_i$ is the length of the $i$-th cloudlet, in MI; $mips_{\alpha}$ is the processing power of the underlying guest entity (configured with virtualization configuration $\alpha$), in MIPS; and $O_{\alpha}$ is the associated virtualization overhead, in seconds. The symbol $bw_{\alpha}$ is the network bandwidth allocated to the guest entity, which is equivalent to 1 gigabit for sending and receiving packets. The network bandwidth is the same across each virtualization configuration, as determined by the components of the simulated datacenter in \Cref{subfig:dc}. The packets exchanged between nodes include either a negligible payload of 1 byte, or a non-negligible payload of 1 gigabyte of data.  The number of network hops required to realize the communication $networkHops$ depends on the cloudlet placement: 0 hops for Configuration I, 1 for Configuration II, and 2 for Configuration III. 

Other factors can degrade the processing power and the average network bandwidth of VMs, containers, and switches, such as: i) the interference of co-located cloudlets; ii) components with heterogeneous hardware capabilities; and iv) more complex network and DAG topologies. A detailed analysis is necessary in such cases to determine the worst-case execution time for each cloudlet and appropriately ``inflate'' the expected makespan of the activation~\cite{andreoli2023rtCloud}. However, this is out-of-scope, hence why for such a simple scenario the makespan estimated with \Cref{eq:makespan} is good enough.

\begin{figure*}
    \centering
    \includegraphics[width=\linewidth]{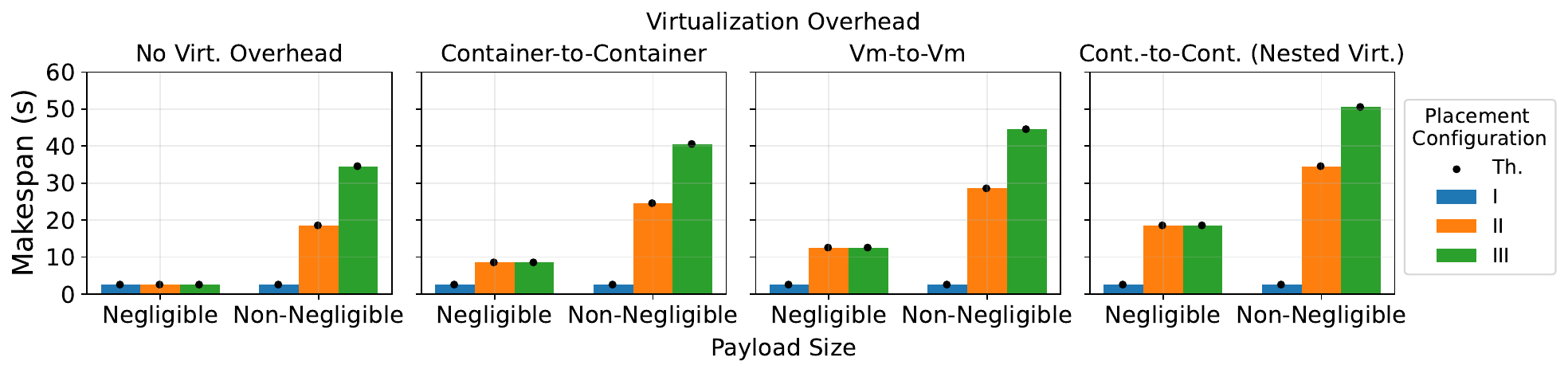}
    \caption{Makespan of a single activation of the workflow application. The black dots correspond to the theoretical makespan estimated with \Cref{eq:makespan}.}
    \label{fig:single-makespan-plot}
\end{figure*}

\begin{figure*}
    \centering
    \includegraphics[width=\linewidth]{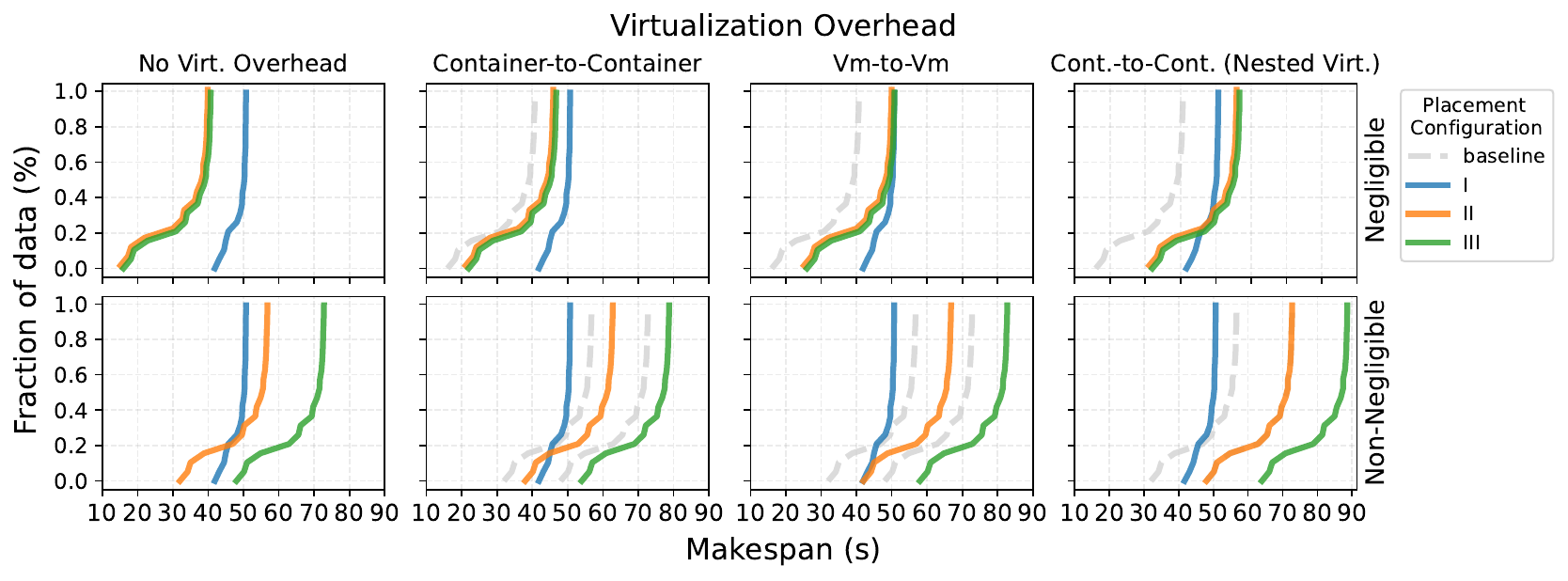}
    \caption{Empirical Cumulative Distribution Function of the makespan for 20 activations of the workflow application. The first row depicts the scenarios with negligible payload size, the second row corresponds to the scenario with non-negligible payload size. The dashed lines highlight the latency shift compared to the baseline scenario (i.e., virtualization overhead disabled).}
    \label{fig:many-makespan-plot}
\end{figure*}
\Cref{fig:single-makespan-plot} illustrates some preliminary experiments where only a single DAG activation is simulated. The left-most plot presents the result for the edge-case where the virtualization overhead feature is disabled. In this scenario, for negligible payload sizes, all placement configurations perform the same since the network delay is insignificant. For a non-negligible payload size, each network hop adds a delay of $\sim$16 seconds. The simulated network delay aligns with the estimation from \Cref{eq:makespan} without virtualization overhead, such that: $\forall \alpha \in \{V,C,N\}: M_{\alpha} = 2.564 + networkHops \cdot 16$. The remaining three plots depict the growing impact of virtualization overhead across the deployments depicted in \Cref{subfig:virt}. The results for negligible payload sizes remain similar to the edge-case for Placement Configuration I, but scaled up for Configuration II and III due to the virtualization overhead. However, there is no distinction between the two configurations that use the network, because the virtualization overhead affects the sending and receiving guests, while physical components like switches remain unaffected. Consequently, for negligible payload sizes, the increase in makespan is solely attributable to the additional overhead introduced by the virtualization deployment, and not by network hops (which remain insignificant), as predicted by the theoretical model.

\Cref{fig:many-makespan-plot} depicts the empirical Cumulative Distribution Function (eCDF) of the makespan for a series of DAG activations. In particular, the DAG is activated 20 times with exponential inter-arrival times so that the computational noise increases incrementally due to the number of concurrent activations. The first row represents scenarios using a negligible payload size, whereas the second one depicts scenarios using a non-negligible payload size. Each column indicates a different virtualization deployment, including the no overhead edge-case.

Placement Configuration I demonstrates consistent behavior across all scenarios, since the network is not used, thus rendering each virtualization deployment identical.
Moreover, in the first column of plots, there is no difference between Placement Configuration II and III due to the negligible payload (as previously depicted in~\Cref{fig:single-makespan-plot}). Hence why, for presentation purposes only, the orange curve has been slightly shifted upwards to eliminate the overlap. Placement Configuration I shows significant computational noise compared to the other configurations for the edge-case scenario. More specifically, its median value is 25\% higher than those of II and III for the no overhead edge-case (top-left plot). This occurs because all cloudlets reside on the same VM: the exponential inter-arrival time of user requests triggers increasingly overlapping activations of both the source and sink nodes of the DAG, leading to high contention. Consequently, the VM becomes progressively congested due to the shared processing power.  On the other hand, distributing the workload across two separate VMs significantly reduces the pressure, resulting in less contention and therefore shorter makespans, even with the exponential inter-arrival time. However, this is gradually overshadowed by the virtualization overhead, as evident when examining the plots from left to right in both the first and second rows. 

The second row of plots shows the result for a non-negligible payload size. Configuration II and III exhibit similar behavior as before, but the curves are shifted further to the right and separated from each other, showcasing the higher makespans of Configuration III due to the additional network hop. In this scenario, Placement Configuration I is the optimal choice for achieving a minimal makespan.

These experiments give some leverage to the Cloud provider in estimating an end-to-end deadline to minimize the number of missed deadlines. For instance, a deadline of 90 seconds would likely guarantee no misses in every possible scenario, though it might be too lax depending on the specific context. A deadline of 50 seconds is optimal only with Placement Configuration I, however, the configuration presents shortcomings in terms of fault-tolerance (i.e., single-point-of-failure) and performance (i.e., it cannot admit a high throughput of concurrent DAG activations due to co-location interferences).

\section{Conclusions and Future Outlook}\label{sec:conclusions}
This article unveils a revival of the CloudSim project~\cite{calheiros2011cloudsim} with the release of its seventh iteration. CloudSim 7G is the release with the most internal changes to date: it offers a novel base layer comprising multiple past modules that have undergone a massive refactoring, optimization and refinement process to accommodate the proposed design changes. Module developers can now integrate, and customize as needed, the essential building blocks of CloudSim, such as physical hosts, VMs and scheduling policies~\cite{calheiros2011cloudsim}, containers~\cite{piraghaj2017containercloudsim}, power-awareness~\cite{beloglazov2013dynamicConsolidation} and simple network modeling~\cite{garg2011networkcloudsim}, within the same simulated scenario. Therefore, CloudSim 7G facilitates the integration of multiple modules, opening new opportunities for the evaluation of scheduling and resource management techniques for next-generation Cloud Computing environments. Furthermore, we have empirically demonstrated the improved performance of CloudSim 7G thanks to the aforementioned refactoring and refinement process. Thanks to our recent optimization effort, CloudSim 7G uses up to 25\% less total memory allocated in selected experiments and overall saves 4-10 seconds compared to CloudSim 6G.

In what follows, we describe a series of future courses of action that we hope will improve CloudSim with the help of the research community. A number of researchers~\cite{Sukhoroslov2022dslab, silva2017cloudsimplus, pucher2015cloudsimAccuracy} have criticized the cloudlet scheduling component accuracy for evaluating production-quality Cloud components.  In this regard, CloudSim 7G represents a significant advancement for accelerating the development of new extensions, including one offering a comprehensive overhaul of the scheduling component so to replace the overly simplistic default models. However, there is a need in this area to support more realistic time-shared CPU scheduling policies for virtual cores of multi-core VMs and parallel processes. Moreover, the network simulation components should be further evolved and extended, in order to provide a richer simulation of the network behavior within distributed Cloud infrastructures.

A further development step of CloudSim's internals concerns the merging of the host and guest entity components into one standardized interface. Indeed, the two concepts realize the same activities, but at different layers: a host entity resides on the ``physical'' layer, whereas a guest entity resides on the ``virtual'' layer. Such a paradigm shift is currently too catastrophic for restoring compatibility with older modules. In this regard, CloudSim 7G serves as an intermediate step toward the realization of a fully generic physical/virtual entity component in future releases.
Secondly, CloudSim might be updated to specify the computational capabilities of hosts and virtual machines in terms of clock frequency, in addition to the current MIPS performance metric. Lastly, CloudSim requires a more comprehensive test suite to ensure its correctness and prevent software regressions between releases.

Given the large research work on approaches based on Machine Learning (ML), Artificial Intelligence (AI) and Reinforcement Learning (RL)~\cite{Lanciano2023}, it may be useful to explore the possibility of integrating within CloudSim additional components to train and evaluate AI/ML models, to simulate AI-enhanced resource management strategies in Cloud Computing. This will require the realization of new interfaces to interact with ML/AI accelerators and development frameworks.

To conclude, we encourage the research community to develop new modules that take advantage of the design shift introduced with CloudSim 7G and to investigate novel scheduling policies with hybrid scenarios using network modeling, power, serverless, NUMA architectures, quantum computing, and VM-container-host placement policies. On a similar note, we kindly request experienced module developers to update their contributions following the guidelines provided in this paper to leverage the novel features of CloudSim 7G.

\vspace{12px}
\noindent \textbf{Software availability} \\ 
\noindent The source code and examples of
the CloudSim7G toolkit are accessible on GitHub\footnote{See: \href{https://github.com/Cloudslab/cloudsim/releases/tag/7.0}{https://github.com/Cloudslab/cloudsim/releases/tag/7.0}} as an open-source tool under the GPL-3.0 license.

\vspace{12px}
\noindent \textbf{Acknowledgments} \\ 
\noindent The authors acknowledge and thank all contributors and developers of CloudSim, and their modules that have been integrated into this new version of CloudSim. This work is partially supported by the Australian Research Council Discovery Project and EU Erasmus+ Traineeship.

\vspace{12px}
\noindent \textbf{Conflict of interest statement} \\ 
The authors declare that they have no known competing financial interests or personal relationships that could have appeared to influence the work reported in this paper.

\bibliographystyle{abbrv}
\bibliography{main.bib}

\end{document}